# Improved performance of traveling wave directional coupler modulator based on electro-optic polymer

Xingyu Zhang*[a], Beomsuk Lee[a], Che-yun Lin[a], Alan X. Wang[b], Amir Hosseini[c], Xiaohui Lin[a], and Ray T. Chen[a]

[a] The University of Texas at Austin, 10100 Burnet Rd, MER160, Austin, TX USA, 78758; [b] Oregon State University, 3097 Kelley Engineering Center, Corvallis, OR USA, 97331; [c] Omega Optics, Inc., 10306 Sausalito Drive, Austin, TX USA, 78759

## ABSTRACT

Polymer based electro-optic modulators have shown great potentials in high frequency analog optical links. Existing commercial LiNibO3 Mach-Zehnder modulators have intrinsic drawbacks in linearity to provide high fidelity communication. In this paper, we present the design, fabrication and characterization of a traveling wave directional coupler modulator based on electro-optic polymer, which is able to provide high linearity, high speed, and low optical insertion loss. A silver ground electrode is used to reduce waveguide sidewall roughness due to the scattering of UV light in photolithography process in addition to suppressing the RF loss. A 1×2 multi-mode interference 3dB-splitter, a photobleached refractive index taper and a quasi-vertical taper are used to reduce the optical insertion loss of the device. The symmetric waveguide structure of the MMI-fed directional coupler is intrinsically bias-free, and the modulation is obtained at the 3-dB point regardless of the ambient temperature. By achieving low RF loss, characteristic impedance matching with 50Ω load, and excellent velocity matching between the RF wave and the optical wave, a travelling wave electrode is designed to function up to 62.5GHz. Domain-inversion poling with push-pull configuration is applied using alternating pulses on a 2-section directional-coupler to achieve a spurious free dynamic range of 110dB/Hz$^{2/3}$. The 3-dB electrical bandwidth of device is measured to be 10GHz.

**Keywords:** electro-optic polymer, directional coupler, traveling wave modulator, domain inversion, spurious free dynamic range

## 1. INTRODUCTION

Polymer based electro-optic modulators have shown great potentials for a variety of applications, such as telecommunication, analog-to-digital conversion, phased-array radar, and electrical-to-optical signal transduction. Existing commercial LiNibO3 Mach-Zehnder (MZ) modulators have intrinsic drawbacks in linearity to provide high fidelity communication. The spurious free dynamic range (SFDR) of high frequency analog optical links is limited by the system noise and the nonlinearity of the modulation process. There have been various efforts to improve the linearity of modulators either electronically [1, 2] or optically [3-5]. A common goal of these efforts is to lower the third-order intermodulation distortions (IMD3s), which appear when multiple tones of signals are carried over a link. A shortcoming common in all these linearization techniques is that the improved linearity is achieved at the expense of simple device design. A Y-fed directional coupler (YFDC) modulator, on the other hand, had been proven to possess a highly linear transfer function without loss of the simplicity of device design [6]. Even higher linearity is achievable when YFDC is incorporated with the $\Delta\beta$-reversal technique [7, 8]. We previously demonstrated a two-section YFDC modulator with $\Delta\beta$-reversal at low modulation frequency as a proof of concept [9], where we achieved the SFDR of 119dB/Hz$^{2/3}$ with 11dB enhancement over the conventional MZ modulator.

In this paper, we demonstrate an electro-optic (EO) polymer based traveling wave directional coupler modulator with $\Delta\beta$-reversal to extend our proven concept to the GHz frequency regime. A traveling wave electrode with a unique design for RF microprobe coupling is fabricated with low RF loss and excellent velocity matching. The bandwidth-length product of 125GHz cm and the 3-dB electrical bandwidth of 10GHz are achieved. The SFDR of $110\pm3$dB/Hz$^{2/3}$ is measured over the modulation frequency of 2–8GHz. In addition, a $1\times2$ multi-mode interference (MMI) 3dB-splitter, a photo-beached refractive index taper, and a quasi-vertical taper are proposed to reduce the optical insertion loss of the device.

*xzhang@utexas.edu; phone 1 512 471-4349; fax 1 512 471-8575

## 2. DESIGN

### 2.1 Polymer optical waveguide design

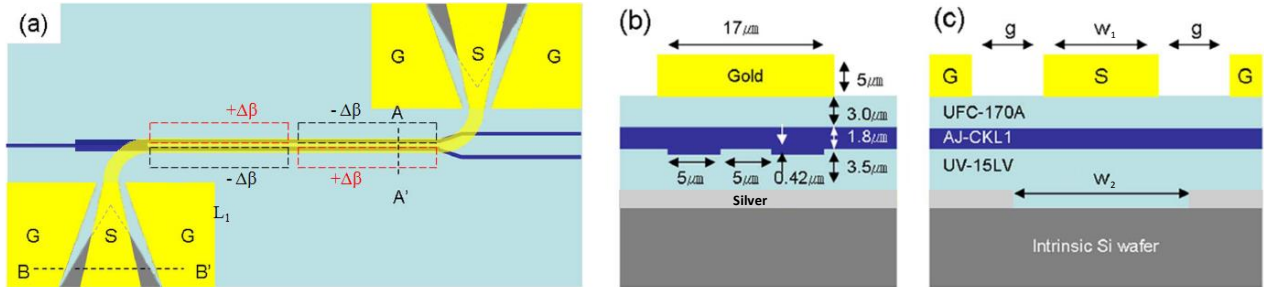

Figure 1. (a) Schematic top view of traveling wave MMI-fed directional coupler modulator with 2-domain-inversion poling, (b) Cross section corresponding to A-A′ in (a), (c) Cross section corresponding to B-B′ in (a). (S: signal electrode, G: ground electrode).

Figure 1 (a) shows the schematic top view of our traveling wave directional coupler modulator. A $1\times2$ MMI 3-dB coupler is designed to equally split the input optical power among two waveguides of a directional coupler as shown in Figure 2 (a). The symmetric waveguide structure of the MMI-fed directional coupler is intrinsically bias-free; and the modulation is automatically set at 3-dB operation point regardless of the ambient temperature. The dimensions of MMI coupler and the optical power distribution in it are shown in Figure 2 (a) and (b), and the total power transmission of this MMI coupler is numerically calculated by FIMMWAVE to be as high as 94%. This MMI coupler has a large fabrication tolerance and is insensitive to the photolithography resolution. On the contrary, as for previously used Y-junction [6], in practice the fabrication limitations in photolithography and etch resolutions usually lead to a blunt tip under a certain distance between the two waveguides (Figure 2 (c) and violates the adiabatic requirement, resulting in extra optical loss [10]. In addition, compared with the previous 1000 μm-long Y-junction [9], the MMI coupler is only 176.6 μm-long and is beneficial to decrease the device length. The inset of Figure 2 (c) shows the fabricated MMI coupler compared with the traditional Y-junction.

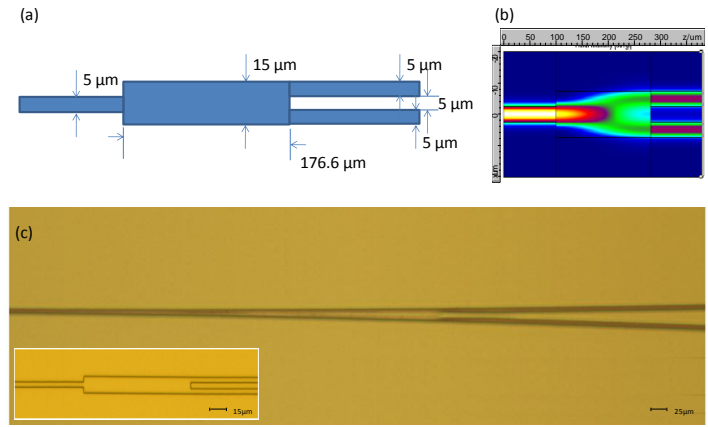

Figure 2. (a) Dimension of $1\times2$ MMI 3-dB coupler, (b) Optical power distribution in the MMI coupler in (a), (c) a blunt tip of a fabricated Y junction due to the fabrication imperfection, and a fabricated MMI coupler shown in the inset.

Unlike the sine-squared transfer curve of the conventional MZ structure, a proper design of coupling length of directional coupler can provide a linear transfer function [1-5]. The linearity of this MMI-fed directional coupler can be further improved to suppress IMD3 by the $\Delta\beta$-reversal technique [6-9]. Multiple-domain-inversion, which helps increase the linearity of directional coupler modulator, has been theoretical demonstrated [11]. Considering the fabrication and poling complexity, in this paper we use a 2-domain-inversion directional coupler for demonstration. The directional coupler is divided into two domains and EO polymer in the first domain is poled in the opposite direction with respect to that in the second domain. Then, the push-pull configuration is also applied to double the EO effect so that the two arms of the directional coupler in each domain are poled in opposite directions. Finally, a single uniform modulation electric field applied by a traveling wave electrode can create $\Delta\beta$-reversal, as shown in Figure 1 (a).

The IMD3 suppression of a directional coupler modulator is a sensitive function of the normalized interaction length ($S_i$), which is defined as the ratio of the interaction length ($L_i$) of i$^{th}$ section to the coupling length ($l_c$). Relative IMD3 suppression of a two-section MMI-fed directional coupler modulator can be graphically represented by plotting the calculated IMD3 suppressions on ($S_1$, $S_2$) plane [12]. $S_1 = S_2 = 2.86$ provides excellent linearity as well as very high modulation depth [11] and is chosen for demonstration in this paper. The coupling length of this directional coupler is matched to 3496 μm for TM mode by tuning the waveguide parameters (FIMMWAVE calculations). The cross section dimensions of this polymer based directional coupler waveguide are shown in Figure 1 (b), and the corresponding total interaction length ($L_1 + L_2$) is 2cm. Figure 3 (a) shows the optical mode profile of this polymer trench waveguide calculated by COMSOL Multiphysics, and the optical effective index is calculated to be 1.599 at 1550nm.

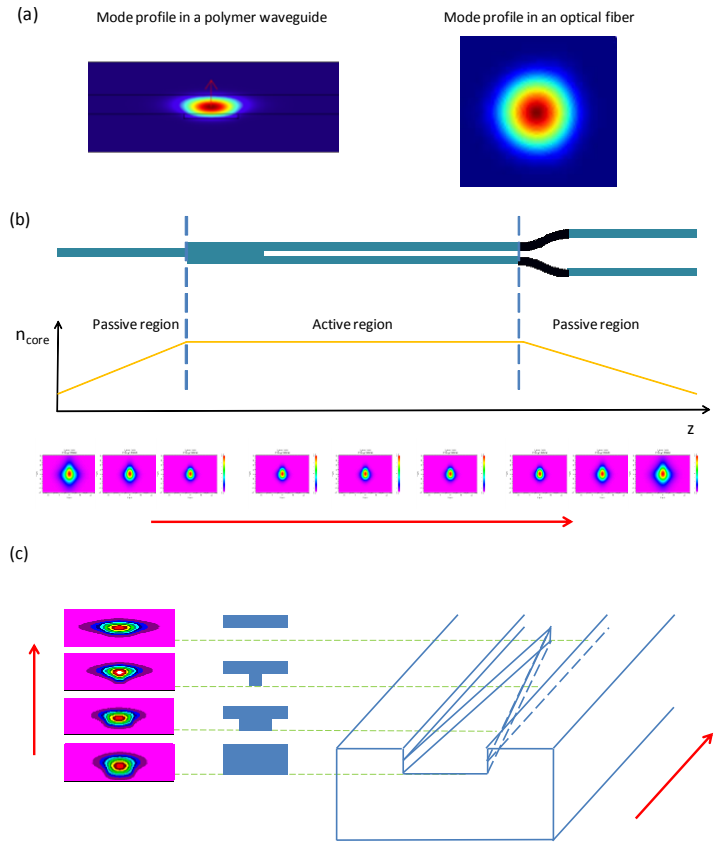

Figure 3. (a) Optical mode profile in a polymer waveguide, compared with the optical mode profile in an input/output single mode fiber. (b) Refractive index taper at passive regions of the MMI-fed directional coupler. The index variation of the photobleached EO polymer in the core layer leads to the gradual change of optical mode size along the taper. (c) Quasi-vertical taper at the facet of polymer waveguide used for mode profile match in vertical direction.

Optical loss is a big concern in the EO polymer modulator design. Figure 3 (a) shows the comparison between the mode profiles of the designed polymer waveguide and that of a single mode fiber both at 1550nm. It can be seen that the mode profile of the polymer waveguide is in an elliptical geometry, with its major axis of about 8μm and minor axis of about 2.5μm, but that the mode profile of the fiber is in a circular geometry with mode field diameter of 10.5±0.8μm. This mode size mismatch can lead to large optical coupling loss. To reduce such coupling loss, a refractive index taper is designed at the passive regions of the waveguides so that the optical mode profile at the input/output (I/O) sides of the polymer waveguide can better match that of the I/O optical fiber. The working mechanism is shown in Figure 3 (b). The refractive index variation along the taper can be created by UV photobleaching method using a gray-scale photomask or discrete step mask-shifting scheme [13]. In addition, considering the large mode size mismatch in vertical direction as shown in Figure 3(a), a quasi-vertical taper structure [14] is designed, as shown in Figure 3(c). The combination of these two tapers can significantly reduce the optical coupling loss. Simulation results by Rsoft show that the coupling loss can be reduced by 50% when the taper is used. In addition to the coupling loss, the roughness of polymer waveguide sidewalls usually causes large propagation loss. Thus, silver is selected as the ground electrode material and its smooth surface helps reduce waveguide sidewall roughness due to scattering of UV light in photolithography process. Compared to other metals, silver is also beneficial to suppress the RF conductor loss owing to its very low resistivity.

## 2.2 Traveling wave electrode design

To extend the operational frequency to the GHz frequency regime, it is essential to design a traveling wave electrode with the velocity matching between RF wave and optical wave as well as the characteristic impedance matching. Considering the alignment of modulation field with the direction of the $\gamma_{33}$ in the poled EO polymer film, which is in vertical direction in our device configuration, a microstrip line, as shown in Figure 1 (b), is a natural choice for the best alignment. In quasi-static analysis, the characteristic impedance $Z_0$ and the microwave effective index $n_m$ can be expressed as [15, 16]

$$Z_0 = \frac{1}{c \ (C \ C_a)} \tag{1}$$

$$n_m = \left(\frac{C}{C_a}\right)^{1/2} \tag{2}$$

where $C_a$ is capacitance per unit length of the electrode structure with the dielectrics replaced by air, $C$ is the capacitance per unit length with the dielectrics present, and $c$ is the speed of light in vacuum. The Frequency-dependent characteristic impedance and microwave effective index can be numerically calculated using HFSS to match 50Ω and optical effective index (~1.599), respectively. Given $\varepsilon_r$=3.2, $h$=8.3 μm and $t$=5 μm from the waveguide dimensions and the fabrication conditions, the characteristic impedance of 50Ω can be matched when w=17 μm. As shown in Figure 4 (a) and (b), over the frequency range of 1–50GHz, the characteristic impedance varies within 49.5–54.5Ω and the microwave effective index varies within 1.56–1.7. The bandwidth-length product due to the velocity mismatch can be calculated as below [17]

$$f \cdot L \cong \frac{c}{4|n_m - n_o|} \tag{3}$$

where $f$ is the modulation frequency, $L$ is the interaction length, $c$ is the speed of light in vacuum, $n_m$ is the microwave effective index of the microstrip line, and $n_o$ is the optical effective refractive index of waveguide. Using Equation 3, the frequency bandwidth can be theoretically calculated to be up to 93GHz for a 2cm-long microstrip line. Figure 4 (c) shows schematic cross section of the designed microstrip line overlaid the contour of the normalized electric potential calculated by COMSOL Multiphysics. It can be seen that both waveguides are under the effect of a uniform modulation field between the microstrip line and the ground electrode and hence the overlap integral between the optical mode and the RF modulation field can be maximized.

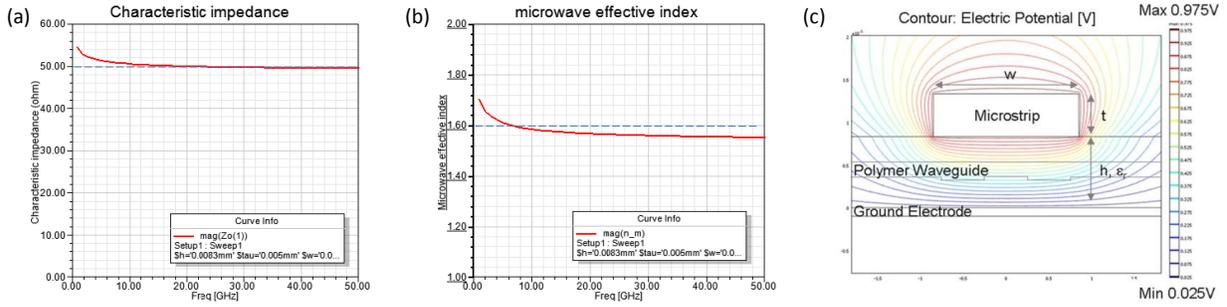

Figure 4. (a) Characteristic impedance of the microstrip line over the frequency range of 1–50GHz calculated by HFSS. The solid red curve indicates the characteristic impedance and the dashed blue line indicates the 50Ω. (b) Microwave effective index of the microstrip line over the frequency range of 1–50GHz calculated by HFSS. The solid red curve indicates the microwave effective index and the dashed blue line indicates the optical effective index (1.599). (c) Schematic cross section of a microstrip line with design parameters overlaid the contour of the normalized electric potential calculated by COMSOL Multiphysics.

To couple the RF power from an GSG microprobe (probe tip width: 50 μm; pitch: 500 μm, as shown in Figure 5 (b)) into the 17 μm-wide microstrip line with minimum coupling loss, a 1.1mm-long quasi-coplanar waveguide (CPW) taper is designed, as shown in Figure 1 (a) and (c). The top width and gap of the coplanar waveguide ($w_1$ and g in Figure 1 (c)) are gradually increased along the taper to match the dimension of a RF microprobe while maintaining the characteristic impedance at 50Ω. In the region of transition from the CPW to microstrip line, the electric field profiles of these two transmission lines should be matched to reduce RF return loss. Therefore, unlike the conventional coplanar waveguide, the ground electrode under the taper is partially removed and the bottom gap ($w_2$ in Figure 1 (c)) is gradually changed along the taper using ground shaping technique [18, 19], based on the consideration that both field matching and impedance matching are important in the transition design. Figure 5 shows the smooth transformation of electric field profile in the CPW-to-microstrip transition as well as 50Ω matching. What is more, for this design to be valid, the resistivity of silicon substrate should be sufficiently high (typical 1 kΩ cm or higher). Otherwise, the finite conductivity of silicon substrate allows the formation of microstrip mode between the signal electrode and silicon substrate, and this microstrip mode would become dominant at wide part of the taper and hinder 50Ω matching.

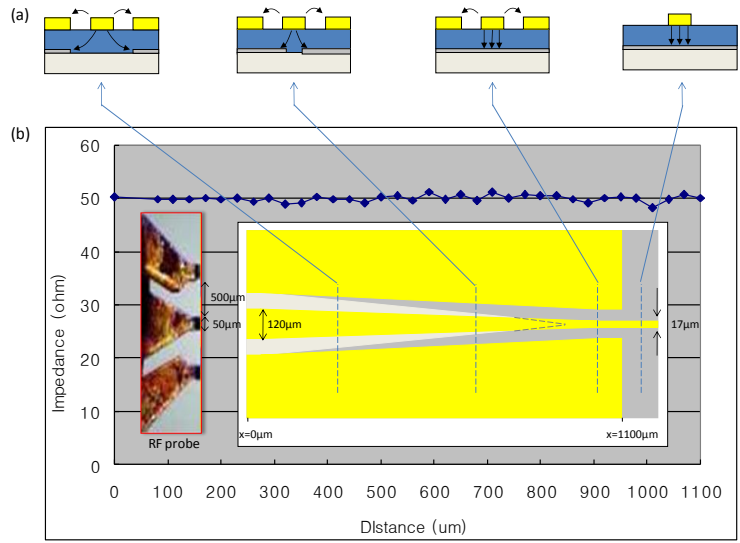

Figure 5. (a) smooth transformation of electric field profile in the CPW-to-microstrip transition. (b) The characteristic impedance (at 10GHz) matched with 50Ω along the taper.

# 3. FABRICATION

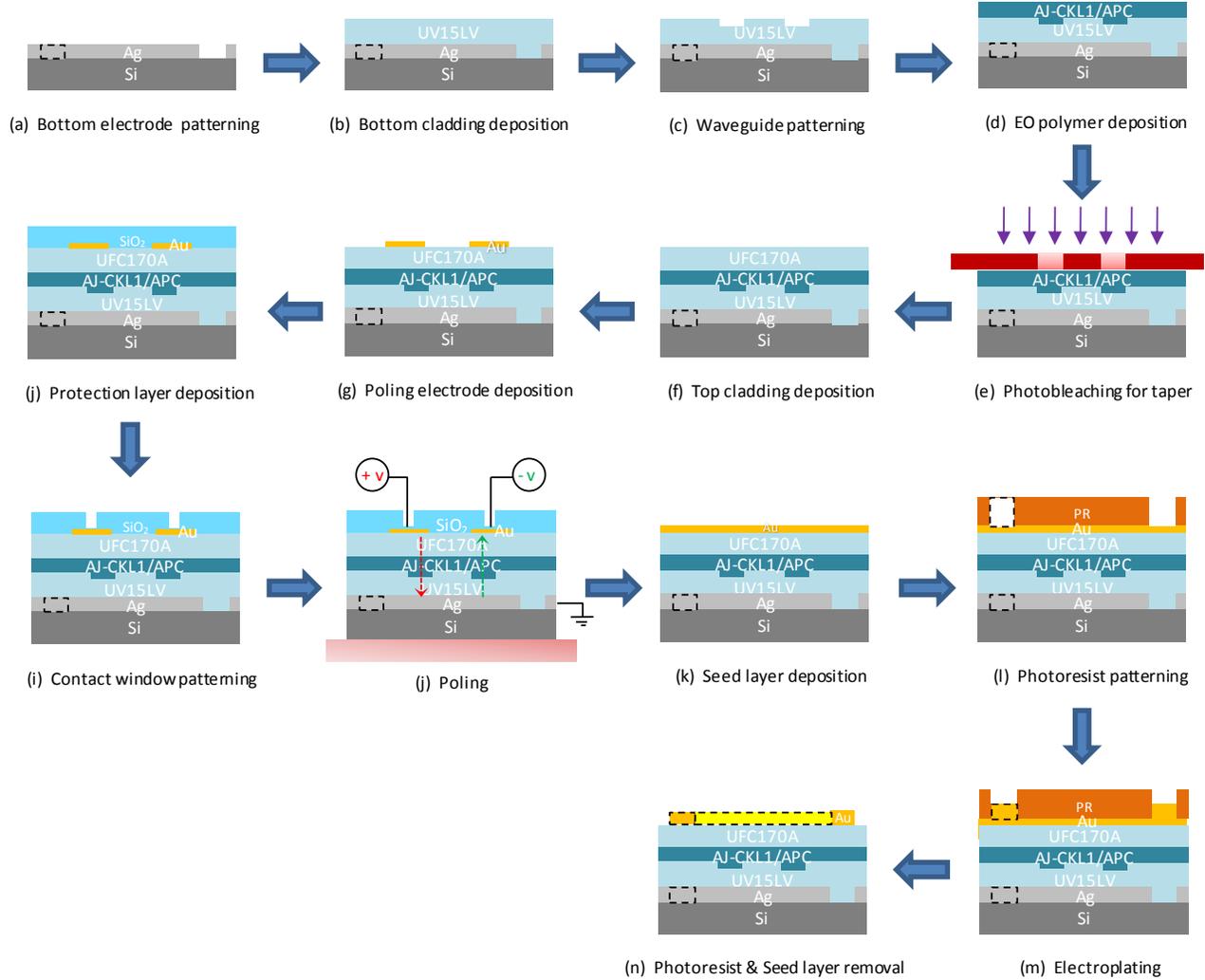

Figure 6. Fabrication process flow

Figure 6 illustrates the fabrication process flow. Device is fabricated on an intrinsic silicon wafer with ultra high resistivity (6–10kΩ cm). A 1 μm-thick silver film is patterned using lift-off process, to serve as the ground electrode for poling process as well as for RF transmission. The polymer trench waveguide consisting of three layers (bottom: UV-15LV, core: AJ-CKL1/APC, and top: UFC-170A, as shown in Figure 1 (b)) is fabricated by photolithography and oxygen plasma reactive ion etching (RIE). The Electro-optic polymer is formulated by doping 25 wt% of AJ-CKL1 chromophore into amorphous polycarbonate (APC). A refractive index taper is fabricated at the passive regions of the waveguides by UV photobleaching using discrete step mask-shifting scheme [13]. For push-pull domain-inversion poling, 150nm-thick gold poling electrodes (indicated by the dashed lines in Figure 1 (a)) are patterned by lift-off process. 300nm-thick silicon dioxide is then deposited above the electrodes and between the gaps of electrodes by e-beam evaporation process. Then, using photolithography and wet etching method, contact windows are opened on the silicon dioxide by wet etching so that the electrodes can be exposed to the probe needles in the following poling process.

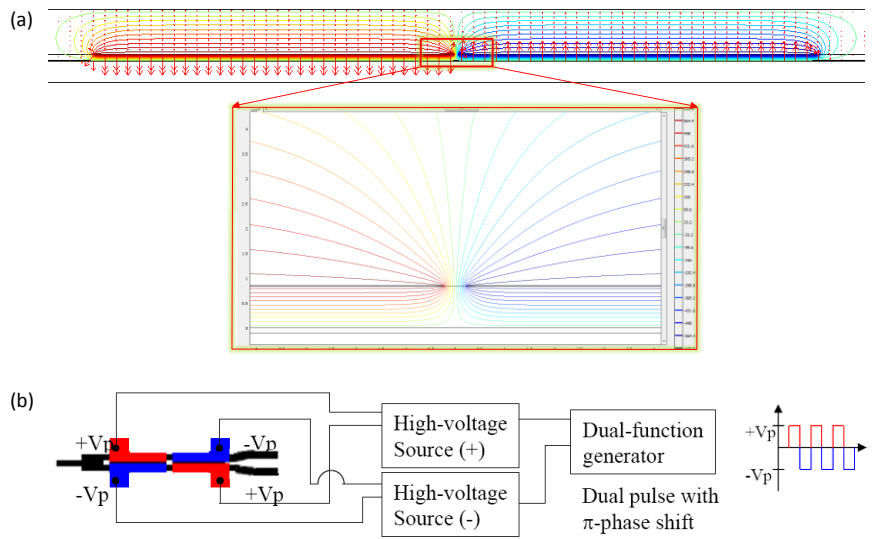

Figure 6. (a) Cross section of poling electrodes above the polymer waveguide overlaid the electric potential distribution in push-pull poling process, calculated by COMSOL Multiphysics. (b) Schematic of push-pull, 2-domain-inversion, alternating-pulse poling.

During the push-pull poling process, two adjacent poling electrodes above the two arms of directional coupler have opposite polarity and the electric field formed between electrodes is very strong, as shown in Figure 6 (a). Given the electrode separation of 5 μm and the polymer waveguide thickness of 8.3 μm, the maximum electric field between two adjacent electrodes can be calculated to be 332V/μm when the electric field of typically ±100V/μm is applied vertically across the polymer waveguide. This increases the probability of dielectric breakdown, normally a big issue in the poling process. To prevent this, the deposited thick silicon dioxide serves as a protection layer due to its good insolating property and high dielectric strength (up to 1000V/μm). According to our testing results, poling can be done with an applied electric field up to 150V/μm at the glass transition temperature ($T_g$=140$^o$C) of EO polymer without dielectric breakdown, which is beneficial to increase poling efficiency. In addition, alternating-pulse poling technique is used to further prevent dielectric breakdown. The positive and negative voltage sources are controlled by dual pulse with π-phase shift from a dual-function generator. Based on the testing experience, the frequency of the alternating pulses should be set to be 1–10Hz to avoid breakdown. During the poling process, the temperature is controlled to increase from room temperature to $T_g$, then maintained at $T_g$ for 5 minutes, and finally quickly decreased back to room temperature. It has been demonstrated that long poling time at $T_g$ can compensate the low poling electric field and increase the poling efficiency [20]. Throughout the entire poling process, poling leakage current is monitored by a picoammeter. Figure 7 shows the leakage current curve depending on the controlled temperature during the poling time.

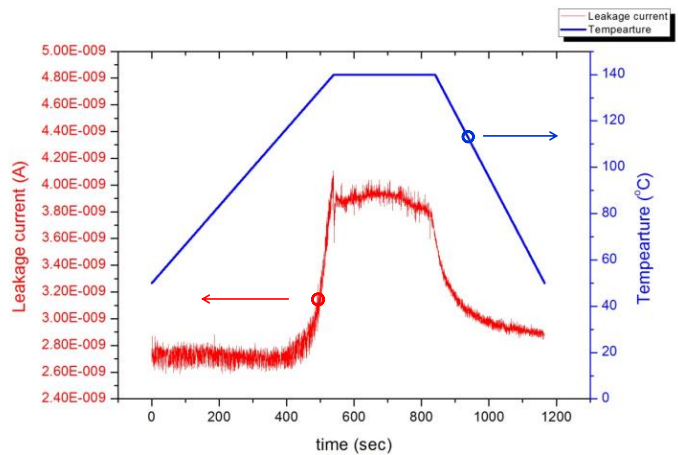

Figure 7. Temperature dependence of leakage current during the poling time.

After poling is done, silicon dioxide and poling electrodes are removed by wet etching method, and finally a traveling wave electrode is fabricated by gold electroplating process. A seed layer of 50nm gold with 5nm chromium adhesion buffer is deposited by e-beam evaporation. The buffer mask is patterned on 10 μm-thick AZ-9260 by photolithography. A 5 μm-thick gold film is electroplated using Techni-Gold 25ES (Technic Inc.). The conductivity of the electroplated gold film is measured to be $2.2 \times 10^7$ S/m. The coplanar and bottom ground electrodes are connected with silver epoxy through via-holes.

## 4. CHARACTERIZATION

### 4.1 RF performance

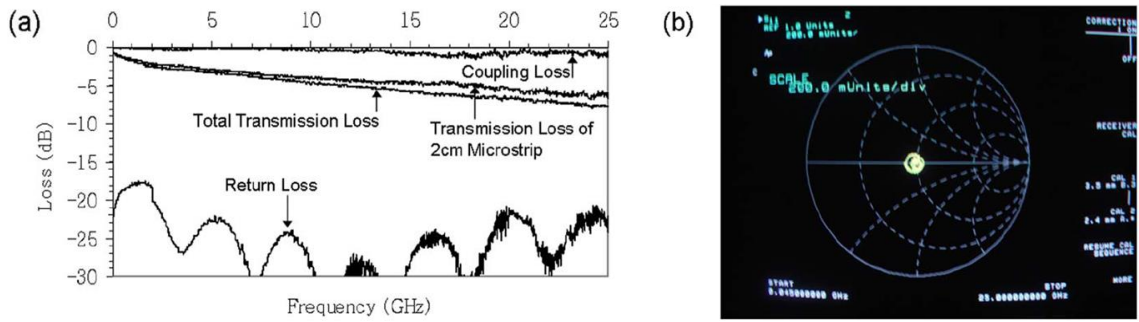

Figure 8. (a) Loss of the fabricated traveling wave electrode. (b) Characteristic impedance of the fabricated traveling wave electrode on Smith chart.

The performance of the fabricated traveling wave electrode is characterized by a vector network analyzer (HP 8510C). An air coplanar probe (ACP40-GSG-250, Cascade Microtech) is used to couple RF power into the tapered quasi-coplanar waveguide. The RF loss of the fabricated traveling wave electrode is presented in Figure 8 (a). The return loss is well below -17dB. The low return loss is probably due to the smooth transformation of electric field profiles in the CPW-microstrp-CPW transition. Square root frequency dependence of transmission loss implies that the RF loss is dominated by the conductor loss, which is measured to be $0.65 \pm 0.05$ dB/cm/GHz$^{1/2}$. Low coupling loss is due to the excellent impedance matching of the tapered quasi-CPW. It is shown in Figure 8 (b) that the characteristic impedance is well centered at 50Ω the Smith chart. The velocity matching between RF wave and optical wave is evaluated by the time domain measurement of the reflection loss, as shown in Figure 9. The effective relative dielectric constant of the microstrip line is measured to be 2.76 and the resulting index mismatch between RF wave and optical wave is 0.06. The bandwidth-length product due to the velocity mismatch can be calculated by Equation (3) to be 125GHz cm, so the modulation frequency limit corresponding to 2cm interaction length would be 62.5 GHz.

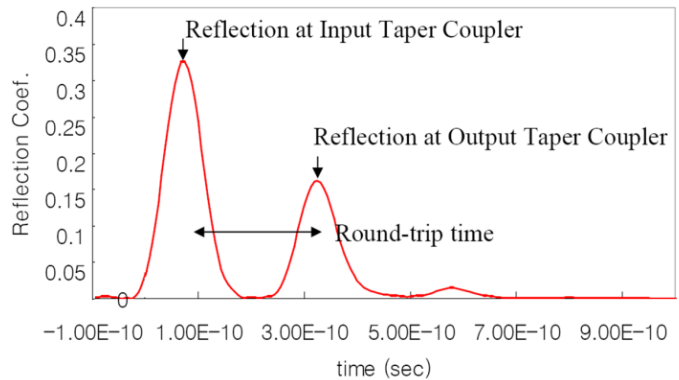

Figure 9. Time domain measurement of reflection loss.

## 4.2 Frequency response

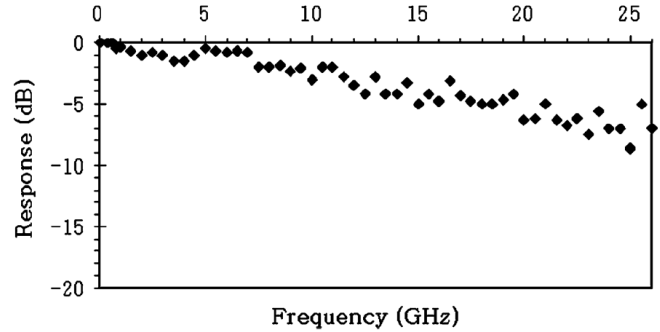

Figure 10. Frequency response of the small signal modulation measured at 4% modulation depth.

The frequency response of the device is evaluated by the small signal optical modulation measured at 4% modulation depth. TM-polarized light with 1.55 m wavelength from a tunable laser (Santec ML-200, Santec Corp.) is launched into the input waveguide through a polarization maintaining fiber and the output light is collected by a single mode fiber. For small signal optical modulation, RF signal from HP 83651B is fed into the traveling wave electrode through the RF microprobe. The modulated optical signal is boosted by an erbium doped fiber amplifier (Intelligain, Bay Spec Inc.), converted to electrical signal by a photodiode (DSC-R409, Discovery Semiconductors Inc.), and then measured by a microwave spectrum analyzer (HP 8560E). The measured 3-dB electrical bandwidth of the device is 10GHz as shown in Figure 10.

## 4.3 Linearity evaluation

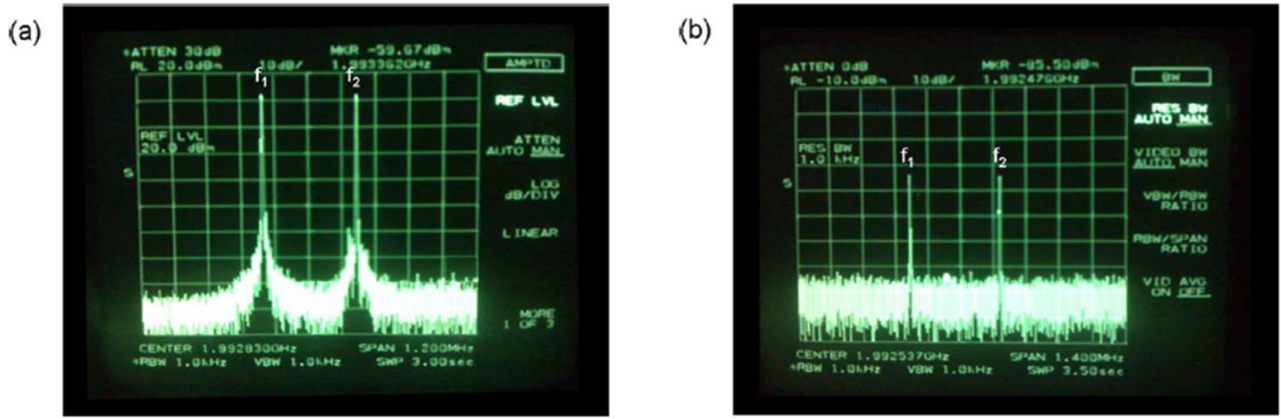

Figure 11. (a) Input two-tone signals ($f_1$ and $f_2$) centered at 1.9928 GHz with 330 kHz tone-interval. (b) Measured output fundamental signals.

A two-tone test is performed to evaluate the linearity of the device. HP 8620C sweep oscillator is used as the second RF source for the two-tone input signals. Agilent 83020A and HP8449B are used as pre- and post-RF amplifier, respectively. The two input RF signals are combined by a coaxial two-way RF power combiner (RFLT2W1G04G, RF-Lambda). New Focus model-1014 is used for optical-to-electrical conversion of the modulated signal. The two-tone input signals and the resulting output signals are shown in Figure 11 (a) and (b), respectively. IMD3 signals, which are supposed to appear at one tone-interval away from the fundamental signals if present, are not observed in Figure 11 (b). A possible reason is that IMD3 signals are well suppressed and buried under the noise floor at this modulation depth. The power level of the two-tone input signals is 12dBm as shown in Figure 11 (a), which is the maximum level available in our two-tone test setup, and this power level translates into the modulation depth of 15%. The simulation result in [11] predicts the IMD3 suppression at 15% modulation depth to be 74dB and the corresponding experimental result in [9] is 69dB, which is a reasonable value considering the fabrication and measurement errors. Neglecting the performance degradation due to RF loss and velocity mismatch, IMD3 signals would be 30dB below the noise floor in Figure 11 (b) at 1kHz bandwidth resolution.

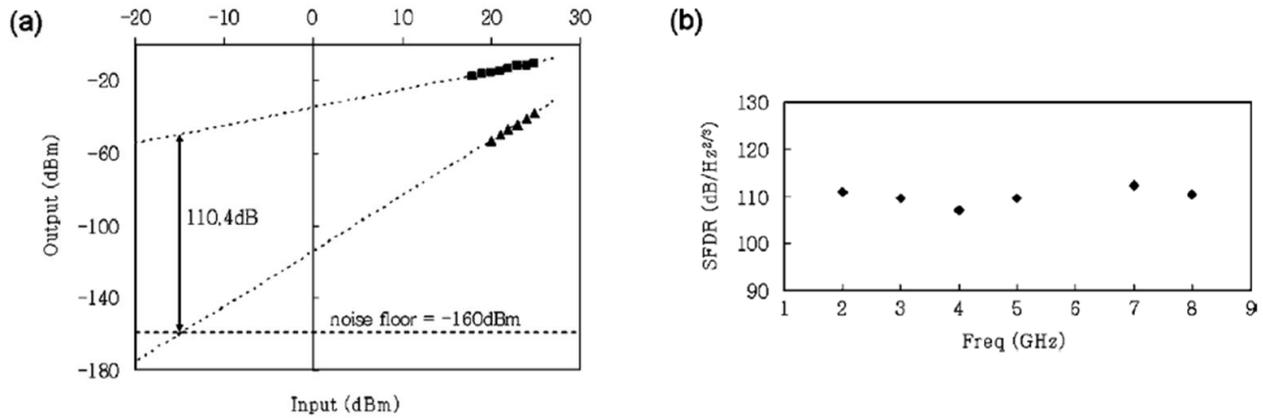

Figure 12. (a) Plot of fundamental and third-order intermodulation distortion signals measured at 8GHz. (b) Spurious free dynamic range measured at 2–8GHz.

Since the IMD3 suppression of the fabricated device is out of the measurable range in our two-tone test setup, SFDR is evaluated through an indirect method. It is known that, with the same modulation depth for both tones, IMD3 is three times or 9.54dB higher than the third harmonic distortion [21]. The power level of mono-tone input signal is extended up to 29dBm by combining the RF source (HP 83651B) with the pre-amplifier (Agilent 83020A). It is found that the third harmonic distortion of our device comes in the detectable range at the mono-tone input signal level above 20dBm. IMD3 signals are obtained by adding 9.54dB to the measured third harmonic distortion signals. SFDR is measured by extrapolating the IMD3 plot to find an intercept point with the noise floor, which is assumed at -160dBm considering the typical fiber-optic link parameters [22], and then measuring the difference with the extrapolated fundamental signal as illustrated in Figure 12 (a). The measured SFDR is within $110\pm3$dB/Hz$^{2/3}$ over the modulation frequency range of 2–8GHz as shown in Figure 12 (b). The low end frequency is determined by the operation range (2–26.5GHz) of the pre-amplifier (Agilent 83020A) and the high end is limited to 8GHz because the third harmonic of the modulation frequency above 8GHz goes beyond the scope (~26.5GHz) of the microwave spectrum analyzer. The SFDR at 6GHz is missing due to the irregular gain of the post-amplifier at 18GHz. Schaffner et al. reported the SFDR of 109.6 dB/Hz$^{2/3}$ at 1 GHz with a lithium niobate directional coupler modulator which is linearized by adding passive bias sections [23]. In their measurement, however, the noise floor was set at -171dBm, which offers 7.3dB extra dynamic range compared with the noise floor at -160dBm. Hung et al. achieved even higher SFDR of 115.5dB/Hz$^{2/3}$ at 3GHz with a linearized polymeric directional coupler modulator by subtracting the distortions of the measurement system [24]. Note that our SFDR of $110\pm3$dB/Hz$^{2/3}$ includes the distortions from the entire measurement system as well as the device.

## ACKNOWLEDGEMENT

Financial supports from the Defense Advanced Research Projects Agency (DARPA) under Contract No. SBIR W31P4Q-08-C-0160 monitored by Dr. Devnand Shenoy is acknowledged.